\documentclass[aps,pre,twocolumn,showpacs]{revtex4}
\usepackage{graphicx}%
\usepackage{dcolumn}
\usepackage{amsmath}   

\usepackage{bm}
\usepackage{longtable}
\begin{document}

\title{ Dynamical transition, hydrophobic interface, and the
  temperature dependence of electrostatic fluctuations in proteins }
\author{David N.\ LeBard}
\author{Dmitry V.\ Matyushov}
\affiliation{Center for Biological Physics, Arizona State University, 
PO Box 871604, Tempe, AZ 85287-1604                   }

\begin{abstract}
  Molecular dynamics simulations have revealed a dramatic increase,
  with increasing temperature, of the amplitude of electrostatic
  fluctuations caused by water at the active site of metalloprotein
  plastocyanin. The increased breadth of electrostatic fluctuations,
  expressed in terms of the reorganization energy of changing the
  redox state of the protein, is related to the formation of the
  hydrophobic protein/water interface allowing large-amplitude
  collective fluctuations of the water density in the protein's first
  solvation shell. On the top of the monotonic increase of the
  reorganization energy with increasing temperature, we have observed
  a spike at 220 K also accompanied by a significant slowing of the
  exponential collective Stokes shift dynamics. In contrast to the
  local density fluctuations of the hydration-shell waters, these
  spikes might be related to the global property of the water solvent
  crossing the Widom line.
\end{abstract}
\pacs{87.14.E-, 87.15.N-, 87.15.Pc }

\maketitle

\section{Introduction}
\label{sec:1}
Dynamical transition has been observed in many hydrated biopolymers,
including proteins, DNA, and RNA
\cite{Rasmussen:92,LeeWand:01,Parak:03,Caliskan:06}. In amounts to a
sharp change in the temperature slope of mean-squared atomic
displacements of the biopolymer atoms at the temperature usually
observed in the range $T_{\text{tr}}=200-230$ K. While the microscopic
origin of this dynamical transition is still debated
\cite{ChenPNAS:06,Kumar:06,Pawlus:08,Khodadadi:08,Ngai:08}, an
important open question is how the existence of this universal
property of hydrated biopolymers \cite{Ringe:03} affects their
physiological activity
\cite{Parak:80,Rasmussen:92,LeeWand:01,Ringe:03}.

Electrostatics is significant to the catalytic action of
enzymes \cite{Warshel:01}.  Therefore, a link between protein's
dynamical transition and enzymatic activity may exist in some property
characterizing electrostatics at the active site.  It is currently
well established that the dynamical transition is not observed in dry
proteins, and its existence is universally attributed to the
interaction of water with the protein interface. A property sensitive
to the dynamical transition needs to connect water's electrostatics to
protein's active site.  Here, we consider one such parameter which
critically affects barriers of protein redox reactions, the
reorganization energy of electron transfer \cite{MarcusSutin}.

The reorganization energy $\lambda$ of electronic transitions between
proteins characterizes the breadth of thermal fluctuations of the
energy gap $\Delta E$ between the donor and acceptor energy levels
\begin{equation}
  \label{eq:1}
  \lambda = \beta \langle \left( \delta \Delta E \right)^2 \rangle/2.  
\end{equation}
Here, $\delta \Delta E$ is the fluctuation of the energy gap $\Delta E$ and the
inverse temperature $\beta=1/(k_{\text{B}}T)$ corrects for the
proportionality of the variance to temperature following from the
fluctuation-dissipation theorem \cite{Landau5}. The reorganization
energy $\lambda$ is then typically a weak function of temperature when
measured for electronic transitions in molecular polar solvents
\cite{DMacc:07}. 

Experimentally accessible reorganization energy of interprotein
electron transfer \cite{Gray:05} characterizes the coupling of the
energy levels of both the donor and acceptor to the thermal bath. For
long-distance electron transfer, most common in biological energy
chains, $\lambda$ can be split into a sum of individual, donor and acceptor,
components and a Coulomb correction. Since these individual components
mostly characterize the physics of the problem, our focus here is on
the electrostatic fluctuations at the active site of a single protein.

Electron transfer changes the redox state of the protein and thus the
partial atomic charges of the active site. The electrostatic
interactions of these charge differences with the potential of the
hydrating water $\phi_{w,j}$ at atomic sites $j$ contribute to the
Coulomb shift $\Delta E_w^C$ which is a part of the overall donor-acceptor
energy gap:
\begin{equation}
  \label{eq:2}
  \Delta E_w^C = \sum_j \Delta q_j \phi_{w,j} .
\end{equation}
Here, the sum runs over the atoms of the active site.  The variance of
this Coulomb energy gap calculated for the charges $\Delta q_j$ of the
active site of a single protein is what is studied in this paper. The
water reorganization energy is then defined as 
\begin{equation}
  \label{eq:10}
    \lambda_w = \beta \langle (\delta \Delta E_w^C)^2\rangle /2 .  
\end{equation}

The dynamical dimension of the problem is characterized by the
normalized Stokes shift correlation function \cite{Jimenez:94}
\begin{equation}
  \label{eq:3}
  S_w(t) = \langle \delta \Delta E_w^C(t) \delta \Delta E_w^C(0)\rangle /  \langle (\delta \Delta E_w^C(0))^2 \rangle ,
\end{equation}
where angular brackets denote an ensemble average.  The common form of
$S_w(t)$ is dense polar liquids includes a fast one-particle component
with a Gaussian decay followed by exponential (or stretched
exponential) decay describing collective solvent dynamics
\cite{Jimenez:94}
\begin{equation}
  \label{eq:4}
  S_w(t) = A_G e^{-(t/ \tau_G)^2} + (1-A_G) e^{-(t/ \tau_E)^{\beta_E}} .
\end{equation}
Here, $\tau_G$ and $\tau_E$ are, respectively, the Gaussian and exponential
relaxation times and $A_G$ quantifies the relative weight of
single-particle dynamics in the reorganization energy; $\beta_E$ is the
stretching exponent.

The main result of the Molecular Dynamics (MD) simulations presented
here is to show that the reorganization energy $\lambda_w(T)$ rises
significantly with temperature to a value much exceeding both the
common estimates of this parameter for reactions involving small redox
molecules,\cite{MarcusSutin} and previous estimates for protein
electron transfer \cite{Warshel:01}. We associate this increase with
the formation of the hydrophobic interface allowing large-amplitude
fluctuations of the local water density. We also show that both the
long-time exponential relaxation time [$\tau_E$ in Eq.\ (\ref{eq:4})] and
the collective part of the reorganization energy $\lambda_E = (1- A_G)\lambda_w$
pass through peaks at the temperature of dynamical transition
$T_{\text{tr}}=220$ K.  This special temperature (at atmospheric
pressure) has been previously associated with a thermodynamic
singularity in the phase diagram of bulk water
\cite{Xu:05,AngellScience:08}. It has also been recently suggested
that a transition from fragile to strong dynamics of hydrated
biopolymers occurs at the same temperature
\cite{ChenPNAS:06,Chen:06,Lagi:08}.

\begin{figure}
  \centering
  \includegraphics*[width=6.5cm]{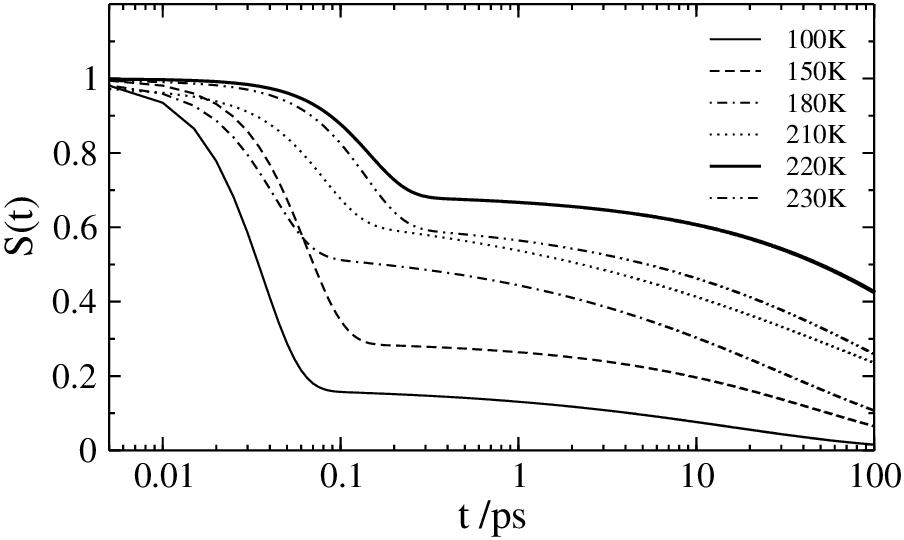}
  \caption{Stokes shift correlation function of PC (Ox) at different temperatures indicated in the plot}
  \label{fig:1}
\end{figure}

\section{MD Simulations}
\label{sec:2}
MD simulations reported here have been done for the redox
metalloprotein plastocyanin (PC) from spinach according to the
simulation protocol described in our previous publication
\cite{DMjpcb1:08}. PC is a single polypeptide chain of 99 residues
forming a $\beta$-sandwich, with a single copper ion ligated by
cysteine, methionine, and two histidines. The protein's active site in
our analysis is composed of a copper ion and four atoms (two
nitrogens and two sulfurs) coordinating it. The partial charges on
these atoms in both reduced (Red) and oxidized (Ox) states can be
found in Ref.\ \onlinecite{DMjcp2:08}.  

The initial configuration of PC was taken from a protonated X-ray
crystal structure with a 1.7 \AA{} resolution (PDB: 1ag6, \cite{1ag698}).
First, the initial protein configuration was minimized in vacuum using
the conjugate gradient method for 10$^4$ steps to remove any bad
contacts.  Then, the system was solvated in an octahedral box with
$N_w = 5886$ TIP3P molecules \cite{tip3p:83}, providing at least two
solvation shells around the protein.  The protein was simulated in the
Ox state with a total charge of $-8$ and in the Red state with the
total charge of $-9$.  In both cases, eight or nine sodium ions were
added to neutralize the system, as is required for the Ewald
summation. After adding the water and counterions, the system's energy
was minimized for another $10^4$ steps while the protein was allowed
to relax and the water and sodium atoms were positionally constrained.
Finally, the entire system was additionally minimized for $10^5$
steps.
 
Following minimization, the system was heated in a $NVT$ ensemble for
30 ps from 0 K to the desired temperature.  Temperature equilibration
was followed by a 2 ns density equilibration in a $NPT$ ensemble at
$P=1$ atm. This equilibrated structure was then used for 20 individual
simulations of the Ox state and 7 simulations of the Red state of PC
to create 10 ns long trajectories.  Temperatures $T$ was varied from
100 to 300 K at constant volume $V$ and constant number of water
molecules $N_w$.  The total simulation time was 324 ns and required
6.9 CPU years, while only 270 ns were used for the production data
analysis which lasted another 2.2 CPU years.  The timestep for all MD
simulations was 2 fs, and SHAKE was used to constrain bonds to
hydrogen atoms. Constant temperature and pressure simulations employed
Berendsen thermostat and barostat, respectively \cite{berend84}.  The
long-range electrostatics were calculated using a smooth particle mesh
Ewald summation with a $9$ \AA{} limit in the direct space sum.

\begin{figure}
  \centering
  \includegraphics*[width=6.5cm]{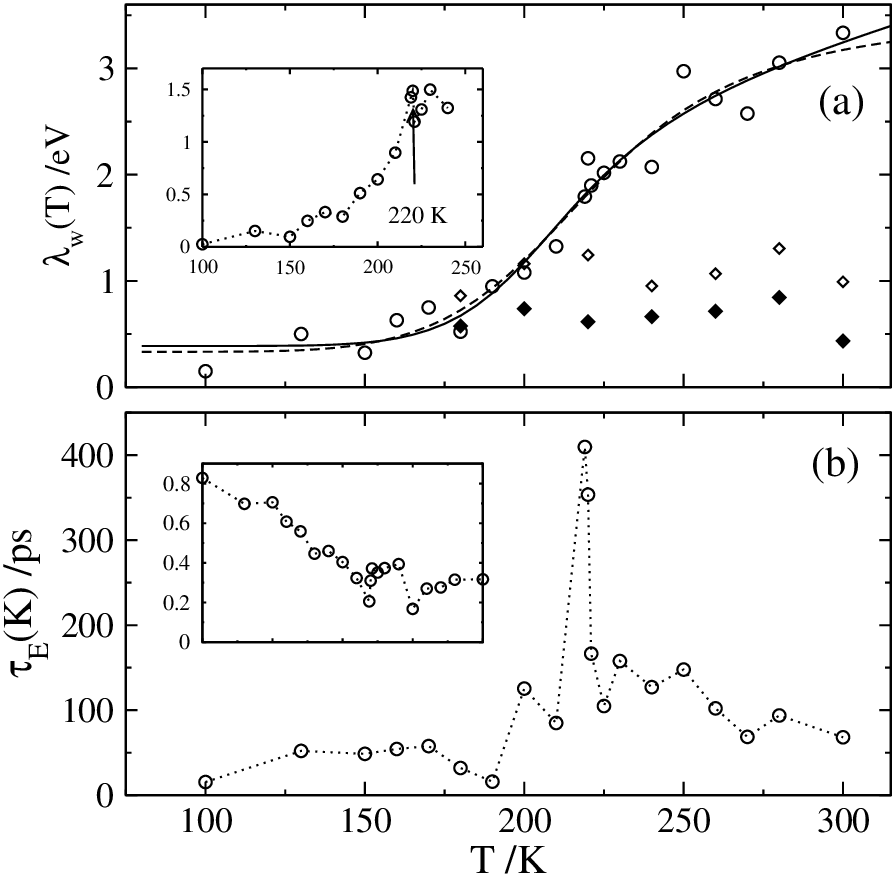}
  \caption{Panel (a): Points are the water reorganization energies
    $\lambda_w$ from MD simulations of PC (Ox) and the solid line
    shows the fit of the simulation data to Eq.\ (\ref{eq:5}) with the
    fitting parameters: $\lambda_G=0.39$ eV, $\lambda_{\text{eq}}=0.87 + 0.0084*T$
    eV, $\tau_{\text{obs}}/ \tau_{0}=7350$, and $E_a=1867$ K; the dashed
    line assumes temperature-independent $\lambda_{\text{eq}}$.  The inset
    shows the exponential part of the reorganization energy related to
    collective water fluctuations. The closed diamonds refer to half
    of the Stokes shift [Eq.\ (\ref{eq:11})] and the open diamonds
    show the linear-response reorganization energy
    $\lambda_w^{\text{Ox}}$ obtained from Eq.\ (\ref{eq:12}).  Panel
    (b): The points refer to the exponential relaxation time of the
    Stokes shift correlation function obtained from MD of PC(Ox) and
    the inset shows the Gaussian amplitude $A_G$ in Eq.\ (\ref{eq:4}).
  }
  \label{fig:2}
\end{figure}

\section{Results}
\label{sec:3}
Most MD results reported here have been obtained from configurations
in equilibrium with the Ox state of PC; the Stokes shift data were
collected from both Ox and Red equilibrium trajectories as discussed
below.  The reorganization energies $\lambda_w(T)$ of PC(Ox) state were
obtained from MD trajectories at different temperatures $T$ and fixed
observation window $\tau_{\text{obs}}=1$ ns. More specifically, the
reorganization energy is calculated from the variance of the Coulomb
energy gap [Eq.\ (\ref{eq:1})] by sliding a 1 ns observation window
along a longer MD trajectory and averaging over the results of the
variance calculations on each window. The average $\langle\Delta
E_w^C\rangle_{\text{obs}}$ required to calculate the variance is not a
global average but is obtained separately from each observation
window. This approach to the calculation of averages is analogous to a
laboratory procedure with a fixed resolution and is required for
studies of systems with broad distributions of relaxation times
\cite{DMjpcb1:08}. In case of proteins, a subset of nuclear motions is
always frozen on the simulation time-scale and so both specifying the
observation window and keeping it constant for all measurements is
significant in maintaining consistent conditions for collecting the
data.

\begin{figure}
  \centering
  \includegraphics*[width=6.5cm]{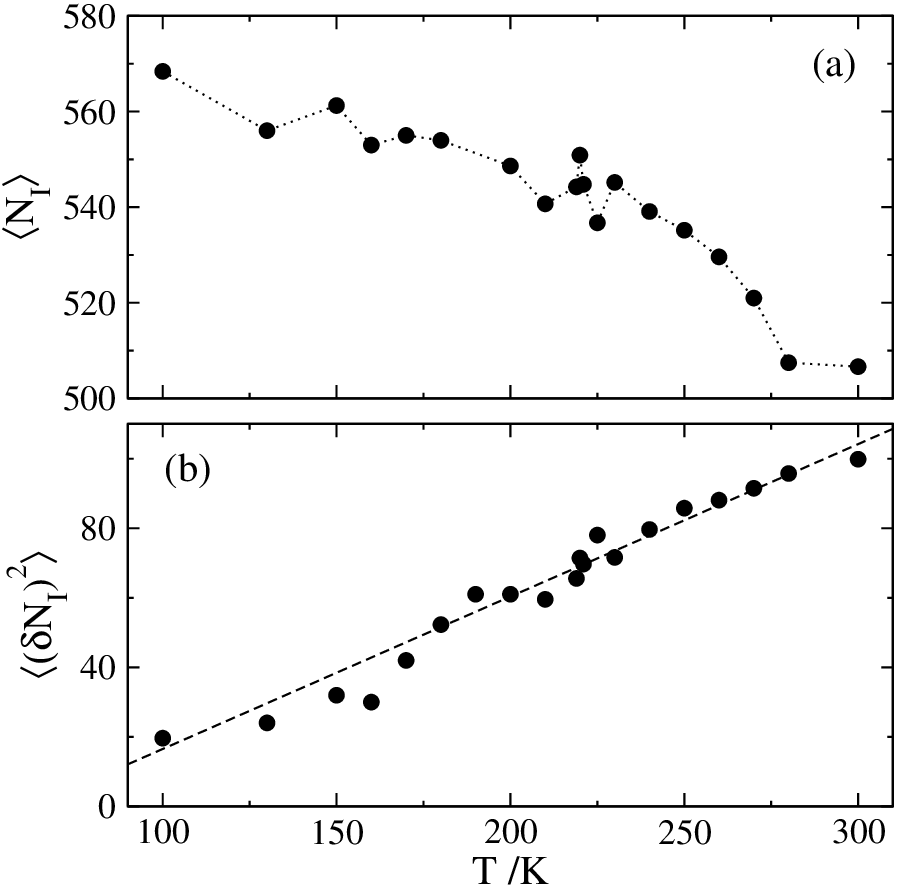}
  \caption{Average number of water molecules in PC's first solvation
    shell (a) and its variance (b) vs temperature.  The dashed line in
    (b) shows a linear regression through the points and the dotted
    line in (a) connects the simulation points. 
 }
  \label{fig:3}
\end{figure}

Fits of the simulated Stokes shift functions to Eq.\ (\ref{eq:4}) are
shown in Fig.\ \ref{fig:1}. Two features are most prominent there: the
increase of the relative importance of the collective solvent dynamics
with increasing temperature (decrease of $A_G$ in Eq.\ (\ref{eq:4})),
and the appearance of a peak in the exponential relaxation time at
$T_{\text{tr}}=220$ K (Fig.\ \ref{fig:2}b). The exponential part of
the reorganization energy $\lambda_E$ also shows a peak at the same
temperature (inset in Fig.\ \ref{fig:2}a).

\begin{table}
  \centering
  \caption{{\label{tab:1}} Properties of hydrated plastocyanin (Ox) from MD
    simulations. }
 \begin{ruledtabular}
  \begin{tabular}{ccccccc}
$T$/K & $\lambda_w$\footnotemark[1] & $\langle N_I \rangle$ & $\langle (\delta N_I )^2 \rangle$ & $p_2$ & $D$\footnotemark[2] & $\tau _E$\footnotemark[3]  \\
\hline
100 & 0.15  & 568 & 20   & $-0.015$ & 0.032  & 15.5  \\
130 & 0.50  & 556 & 24   & $-0.022$ & 0.003  & 52.0  \\
150 & 0.32  & 561 & 32   & $-0.025$ & 0.017  & 48.6  \\
160 & 0.63  & 553 & 30   & $-0.021$ & 0.015  & 54.3  \\
170 & 0.75  & 555 & 42   & $-0.030$ & 0.038  & 57.8  \\
180 & 0.52  & 554 & 52   & $-0.025$ & 0.041  & 32.0  \\
190 & 0.95  & 552 & 61   & $-0.025$ & 0.020  & 15.9  \\
200 & 1.08  & 549 & 61   & $-0.027$ & 0.037  & 125.5  \\
210 & 1.33  & 541 & 60   & $-0.027$ & 0.061  & 84.9  \\
219 & 1.79  & 544 & 66   & $-0.025$ & 0.088  & 409.6  \\
220 & 2.15  & 537 & 71   & $-0.021$ & 0.093  & 353.6  \\
221 & 1.90  & 545 & 70   & $-0.025$ & 0.095  & 166.5  \\
225 & 2.02  & 537 & 78   & $-0.021$ & 0.107  & 104.8  \\
230 & 2.12  & 545 & 72   & $-0.021$ & 0.126  & 157.8  \\
240 & 2.07  & 539 & 80   & $-0.020$ & 0.165  & 127.0  \\
250 & 2.97  & 535 & 86   & $-0.020$ & 0.206  & 147.8  \\
260 & 2.71  & 530 & 88   & $-0.018$ & 0.281  & 102.1  \\
270 & 2.57  & 520 & 91   & $-0.016$ & 0.326  & 68.5 \\
280 & 3.05  & 507 & 96   & $-0.012$ & 0.389  & 93.8 \\
300 & 3.33  & 507 & 100  & $-0.014$ & 0.536  & 68.1 \\ 
\end{tabular}
\end{ruledtabular}
\footnotetext[1]{Water reorganization energies (in eV), obtained with
  $\tau_{\text{obs}}=1$ ns observation window. }
\footnotetext[2]{Diffusion coefficients of TIP3P water averaged over
  all molecules in the simulation box (in \AA$^2$/ps). }
\footnotetext[3]{Exponential relaxation time (in ps) of the Stokes
  shift correlation function in Eq.\ (\ref{eq:4}). }
\end{table}

Overall $\lambda_w(T)$ strongly increases from a value typical for short MD
simulations of proteins \cite{Warshel:01} to a much larger value at
higher temperatures (Fig.\ \ref{fig:2}a and Table \ref{tab:1}). The
temperature of the onset of the $\lambda_w(T)$ rise is much below
$T_{\text{tr}}$, at about 150 K commonly associated with the onset of
rotation of methyl groups of protein's side chains
\cite{Khodadadi:08,Krishnan:08}.  This onset temperature is however
depends on the observation window. Since the relaxation times of the
protein are widely different, the rise of $\lambda_w(T)$ is caused by the
appearance of a particular relaxation mode in the observation window,
methyl rotations in this case. However, we believe that the underlying
picture is more complex and the main rise of $\lambda_w(T)$ is caused not by
methyl rotations, but by a more collective mode coupled to the solvent
interfacial translations \cite{Vitkup:00,Tarek_PhysRevLett:02} (see
below). In fact recent extensive simulations of the mean-squared
atomic displacements of myoglobin \cite{Krishnan:08} have reveled two
breaks in the temperature slope: the first break at 150 K related to
methyl (anharmonic) rotations followed by a stronger solvent-induced
break at 220 K.

The appearance of a relaxation mode in the observation window restores
the statistical ergodicity for that particular mode.  The non-ergodic
rise of $\lambda_w(T)$ to its equilibrium value $\lambda_{\text{eq}}(T)$, also
seen for model charge-transfer chromophores \cite{DMjcp2:06}, can be
described by imposing a step-wise frequency filter on the spectrum of
Stokes shift fluctuations \cite{DMacc:07}
\begin{equation}
  \label{eq:9}
  \lambda_w(T) = 2\lambda_{\text{eq}}(T) \int_{1/ \tau_{\text{obs}}}^{\infty} 
                 S_w(\omega) d\omega .
\end{equation}
Here, $S_w(\omega)$ is the Fourier transform of the Stokes shift
correlation function in Eqs.\ (\ref{eq:3}) and (\ref{eq:4}).  In order
to provide a physically transparent form for $\lambda_w(T)$ one can consider
an effective single-exponential Debye relaxation, instead of several
relaxation modes, to characterize collective nuclear motions coupled
to the Stokes shift dynamics. This procedure leads to the following
simple relation
\begin{equation}
  \label{eq:5}
  \lambda_w(T) = \lambda_G + (\lambda_{\text{eq}}(T) -\lambda_G) (2/ \pi)
  \mathrm{arctg}\left[\tau_{\text{obs}}/ \tau(T) \right] ,
\end{equation}
where the effective Debye relaxation time is given by the Arrhenius
law, $\tau(T)=\tau_0 \exp[\beta E_a]$.

The Gaussian component of the solvent reorganization energy, related
to ballistic water motions \cite{Jimenez:94}, is normally reasonably
temperature-independent \cite{DMjcp2:06}. On the other hand, the
temperature decrease of $A_G(T)\simeq \lambda_G / \lambda_{\text{eq}}(T)$ in the fit of
the Stokes shift function (inset in Fig.\ \ref{fig:2}b) clearly points
to the equilibrium reorganization energy increasing with
temperature. From the anticipated relation of $\lambda_{\text{eq}}(T)$ with
the variance of the number of particles in the first solvation shell,
which linearly grows with temperature (see below), we have attempted a
linear temperature dependence of $\lambda_{\text{eq}}(T)$ to fit the MD data
to Eq.\ (\ref{eq:5}). The result is shown by the solid line in Fig.\
\ref{fig:2}a, and it is not much different from the fit using a
temperature-independent $\lambda_{\text{eq}}$ (dashed line in Fig.\
\ref{fig:2}a). We also note that since our simulation length obviously
cuts some slow nuclear modes off, we have not used $A_G(T)$ from the
fits of the Stokes shift correlation functions to calculate
$\lambda_{\text{eq}}(T)$.

The activation energy $E_a$ of the effective Debye mode obtained from
the fit, $E_a= 1867$ K, points to a secondary $\beta$-relaxation mode
creating fluctuations of the electrostatic potential, in contrast to
the primary $\alpha$-relaxation of the water/protein system with a commonly
much higher activation barrier \cite{Green:94,Fenimore:04}. This
activation energy is also lower than $\beta$ relaxation of aqueous
mixtures with the activation energy of the order of $5.5\times 10^3$ K
\cite{Ngai:08}.

Also shown in Fig.\ \ref{fig:2}a (closed diamonds) is the
reorganization energy from the Stokes shift obtained from the
difference of average Coulomb energy gaps in Ox and Red states:
\begin{equation}
  \label{eq:11}
  \lambda_w^{\text{St}} = \frac{1}{2}\left| \langle \Delta E_w^C \rangle_{\text{Ox}} - \langle \Delta E_w^C \rangle_{\text{Red}} \right|  
\end{equation}
For water fluctuations following linear response one expects the
reorganization energy from the variance [Eq.\ (\ref{eq:10})] to be 
connected to the reorganization energy from the two first moments 
[Eq.\ (\ref{eq:11})] by the following relation:
\begin{equation}
  \label{eq:12}
   \lambda_w^{\text{Ox}} = \lambda_w^{\text{St}} - 
                    (\beta/2)\langle\delta \Delta E_w^C \delta \Delta E_P^C \rangle_{\text{Ox}}. 
\end{equation}
In Eq.\ (\ref{eq:12}) we have stressed that the averages are taken
over the configurations in equilibrium with PC(Ox) and $\Delta E_P^C$ is
the Coulomb interaction energy of the difference charges of the active
site [$\Delta q_j$, Eq.\ (\ref{eq:2})] with the remaining partial charges
of the protein matrix.

When a rigid molecule is solvated and the intramolecular energy gap
does not fluctuate, the second correlator in Eq.\ (\ref{eq:12}) is
zero. One arrives then at the standard expectation of the linear
solvation theories that two routes to the reorganization energy, from
the second cumulant [Eq.\ (\ref{eq:10})] and from two first cumulants
[Eq.\ (\ref{eq:11})], are equivalent
\cite{MarcusSutin,DMacc:07}. Since the protein matrix fluctuates
itself, the cross-correlation in principle needs be taken into
account, and it turns out to be negative \cite{Nilsson:05}. However,
when cross-correlation term is subtracted from $\lambda_w^{\text{St}}$ in
Eq.\ (\ref{eq:12}) (open diamonds in Fig.\ \ref{fig:2}a), the result
is still significantly below the reorganization energy from the
variance [Eq.\ (\ref{eq:10})]. We therefore observe here a severe
breakdown of linear solvation.

What Fig.\ \ref{fig:2}a in fact indicates is that the two definitions
of the reorganization energy converge at low temperatures with the
reorganization energy from the variance deviating significantly upward
above $T \simeq 200$ K.  This observation implies that fast water's modes
coupled to electrostatic fluctuations, presumably librations, which
are still unfrozen at low temperatures, follow the expectations of the
linear response theories. On the contrary, a slower collective mode,
which appears in the observation window at higher temperatures and
gives rise to the gigantic reorganization energy, does not follow the
linear response. The cross-correlation does not restore the linear
response, in contrast to an earlier observation made for a
water-exposed tryptophan residue \cite{Nilsson:05}. The low value of
the cross-correlation physically implies that the elasticities of the
protein and water are drastically different and their electrostatic
fluctuations are mostly decoupled. From that perspective, this
correlation decoupling should hold for any solute/solvent combination
with a significantly different rigidity.

The fluctuations of water's electrostatic potential at the active site
can generally be traced back to two weakly correlated nuclear modes in
polar liquids, the orientational polarization and the local density
\cite{DMmp:93}. In order to clarify the origin of the dramatic rise of
the reorganization energy, we have looked at two additional
correlation functions characterizing the density and orientational
manifolds of the water molecules in the protein's first solvation
shell.  A water molecule is defined as to belong to the first
solvation shell if its oxygen atom is within 2.87 \AA{} distance from the
protein van der Waals surface.

The density manifold is characterized by the fluctuation of the number
of particles $N_I(t)$ in the first solvation shell,
\begin{equation}
  \label{eq:6}
      C_N(t) = \langle \delta N_I(t) \delta N_I(0)\rangle. 
\end{equation}
Further, the orientational manifold is described by the
fluctuations of the total dipole moment $\mathbf{M}(t)$ of the water
dipoles in the first solvation shell, 
\begin{equation}
  \label{eq:8}
  C_M(t) = \langle N_I\rangle^{-1} \langle \delta \mathbf{ M}(t) \cdot \delta \mathbf{M}(0) \rangle,   
\end{equation}
where $N_I(T) = \langle N_I\rangle$ is the average number of waters in the first
solvation shell. In $C_N(t)$ and $C_M(t)$, the fluctuations $\delta N_I(t)$
and $\delta \mathbf{M}(t)$ denote the deviations from the corresponding
average values.  The variances were calculated on the 1 ns observation
window by using the same procedure as for the reorganization energy
calculations.

\begin{figure}
  \centering
  \includegraphics*[width=6.5cm]{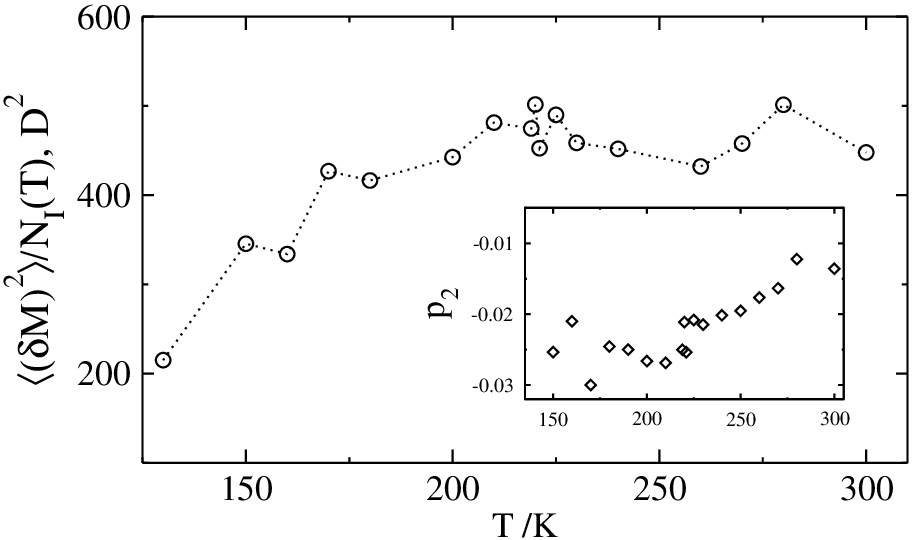}
  \caption{$\langle(\delta M)^2\rangle/N_I(T)$ vs temperature. The inset shows the
    second-rank orientational order parameter $p_2(T)$ [Eq.\
    (\ref{eq:7})]. }
  \label{fig:4}
\end{figure}

The temperature dependences of the average and variance of the number
of waters in the first solvation shell (Fig.\ \ref{fig:3}) are
indicative of the formation of the hydrophobic protein/water interface
with increasing temperature. The average $N_I(T)$ is generally a
decaying function (Fig.\ \ref{fig:3}a), and the slope of this decay
becomes sharper above the transition temperature $T_{\text{tr}}$ (see
below). The decrease in the density of water at the interface allows
stronger density fluctuations (Fig.\ \ref{fig:3}b) and it is this
regime of large interfacial density fluctuations that is a signature
of hydrophobic solvation \cite{ChandlerNature:05}.  In this regime,
one-particle exchanges of water molecules between the surface and the
bulk \cite{Tarek_PhysRevLett:02} combine into large-scale collective
density waves producing significant modulations of the electrostatic
potential reflected in $\lambda_w(T)$. This thermal noise of hydrophobic
surfaces is also reflected in a well-documented increase of protein's
heat capacity upon unfolding, indicative of an increased breadth of the
energy fluctuations \cite{HuangPNAS:00,Prabhu:05}.

The interfacial density fluctuations originate from the exchange of
waters between the hydration shell and the bulk. These fluctuations
can be represented as binding/unbinding events at the protein surface
\cite{DMcp:08} with the resulting equilibrium reorganization energy
$\lambda_{\text{eq}}(T)$ scaling linearly with the variance of the number of
particles in the hydration shell: $\lambda_w(T) = a + b \langle (\delta N_I)^2(T)\rangle $,
where coefficients $a$ and $b$ are weak functions of
temperature. This expectation, used in the solid-line fit in Fig.\
\ref{fig:2}a, is corroborated quite well given the linear scaling of $
\langle (\delta N_I)^2(T)\rangle$ with temperature (Fig.\ \ref{fig:3}b).  A fairly
significant temperature rise of $\lambda_{\text{eq}}(T)$ (see the fitting
parameters in Fig.\ \ref{fig:2}) also indicates a substantial density
component in the overall reorganization energy at ambient conditions,
in contrast to a 20--30\% contribution for small solutes in dense polar
solvents \cite{DMmp:93}. We therefore conclude that the contribution
to $\lambda_{\text{eq}}(T)$ from density fluctuations is significantly
magnified by the soft and flexible nature of the hydrophobic
protein/water interface resulting in a gigantic magnitude of the
overall reorganization energy far exceeding half of the Stokes shift
[Eqs.\ (\ref{eq:11}) and (\ref{eq:12})].

The orientational fluctuations of the first-shell dipoles do not show
a resolvable correlation with the reorganization energy (Fig.\
\ref{fig:4}). The variance of the first-shell dipole moment grows with
rising temperature, in accord with a general expectation of increased
softness of the solvation shell, but does not show an obvious
correlation with $\lambda_w(T)$. There is a weak maximum at $T_{\text{tr}}$
for $\langle (\delta \mathbf{M})^2\rangle$, but it is hard to assess from our data
whether this is another reflection of the same spike seen for the
collective part of the reorganization energy $\lambda_E$ in Fig.\
\ref{fig:2}a.

\begin{figure}
  \centering
  \includegraphics*[width=6.5cm]{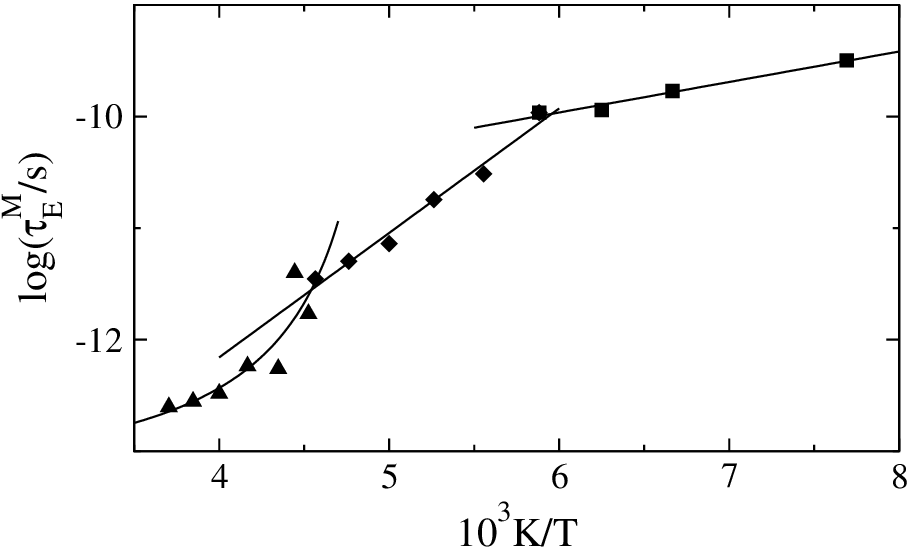}
  \caption{Exponential relaxation time $\tau_{E}^{M}$ extracted by
    fitting the correlation function $C_M(t)$ from Eq.\ (\ref{eq:8}) to
    Eq.\ (\ref{eq:4}). The points are the simulation results in three
    ranges of temperature where they are fitted to the Vogel-Fulcher
    temperature law (at highest temperatures) and to Arrhenius laws (at
    intermediate and lowest temperature ranges). The solid lines show
    the results of the fits.  }
  \label{fig:5}
\end{figure}

The inset in Fig.\ \ref{fig:4} shows the second-rank orientational
order parameter
\begin{equation}
  \label{eq:7}
  p_2 = \left\langle \sum_{j \in I} P_2(\mathbf{\hat e}_j\cdot \mathbf{\hat
      r}_j) \right\rangle .
\end{equation}
Here, $\mathbf{\hat e}_j$ and $\mathbf{\hat r}_j$ are the unit vectors
of the dipole moment and position of molecule $j$ which belongs to the
first solvation shell, and $P_2(x)$ is the second Legendre polynomial.
The low-temperature portion of $p_2(T)$ is practically constant
showing a slight preferential orientation of the water molecules
parallel to the interface. This type of ordering has been previously
observed at interfaces of nonpolar substances and proteins with water
\cite{Lee:84,Gerstein:93}. This preferential ordering decays with
increasing temperature resulting in essentially random, on average,
orientations of water dipoles in the hydration shell. The fairly large
amplitude of the dipole moment fluctuations is therefore most likely
caused by the density fluctuations.

This assessment is supported by the data for exponential relaxation
times of $C_N(t)$ and $C_M(t)$ obtained by fitting these correlation
functions to Eq.\ (\ref{eq:4}). When both exponential relaxation times
are fitted to Arrhenius laws, they produce activation energies of 1389
K and 2076 K, respectively, in a close range with the activation
energy of 1867 K obtained from the fit of $\lambda_w(T)$ to Eq.\
(\ref{eq:5}).  We note that this activation barrier is consistent with
the activation enthalpies of 1400--2400 K obtained by a variety of
techniques for $\beta$-fluctuations of hydrated proteins
\cite{Fenimore:04} which are considered to be slaved by
$\beta$-fluctuations of the hydration shell
\cite{Fenimore:04,Swenson:07}. One also needs to keep in mind that an
average Arrhenius slope actually hides a fairly complex behavior.
Figure \ref{fig:5} shows the exponential relaxation time of $C_M(t)$
vs inverse temperature. The low-temperature portion of the data
(triangles) is well approximated by a non-Arrhenius Vogel-Fulcher
temperature law. This is followed by what can be characterized as a
fragile-to-strong crossover followed by yet another break in the
Arrhenius slope at $\simeq 160$ K. This picture is consistent with two
breaks in the slope seen in the simulations of mean-squared atomic
displacements of myoglobin \cite{Krishnan:08}, where the
lowest-temperature break was associated with the onset of methyl group
rotations. The results for exponential relaxation times of $C_N(t)$
are more scattered and we could not reach an equally informative
conclusion except for the average Arrhenius slope.

\section{Discussion}
\label{sec:4}
Many alternative explanations have been sought for the observed
dynamical transition in biopolymers \cite{Ringe:03}. Given that the
transition is not observed for dry protein samples, the possible
scenarios are limited to either the protein/water interface or to a
bulk property of water. The recent observations, from neutron
scattering measurements, of the fragile-to-strong crossover in the
dynamics of partially hydrated protein powder samples
\cite{ChenPNAS:06} point to the second (bulk water) scenario.  The
crossover, also seen in the recent simulations
\cite{Kumar:06,Lagi:08}, can be connected to the bulk water crossing
the Widom line, i.e.\ the line of maximum cooperativity of the water
fluctuations \cite{Xu:05}. On the other hand, other recent
experimental data on quasielsatic neutron scattering, dielectric
relaxation \cite{Khodadadi:08}, and conductivity \cite{Pawlus:08} of
hydrated proteins have not revealed any special points in the
corresponding relaxation times around the temperature of dynamical
transition. These latter data report the temperature dependence of the
primary $\alpha$-relaxation of the protein/water system and therefore these
authors have concluded that the observed dynamical crossover
\cite{ChenPNAS:06} should be attributed the appearance of a secondary
relaxation in the observation window at $T>T_{\text{tr}}$
\cite{Swenson:07,Khodadadi:08}.

In addition, recent observations of the dynamical transition in DNA
and RNA \cite{Caliskan:05,Chen:06} have clearly shown that this
property is not unique to a peptide-based polymer. These findings
again re-emphasize the notion that either a bulk property of water or
some generic property of the interface, not much sensitive to the
details of the macromolecular structure, are responsible for the
transition. Our data in fact suggest that both bulk and interfacial
views need to be invoked to explain different facets of the problem,
but the interface aspect has a dominant effect.

\begin{figure}
  \centering
  \includegraphics*[width=6.5cm]{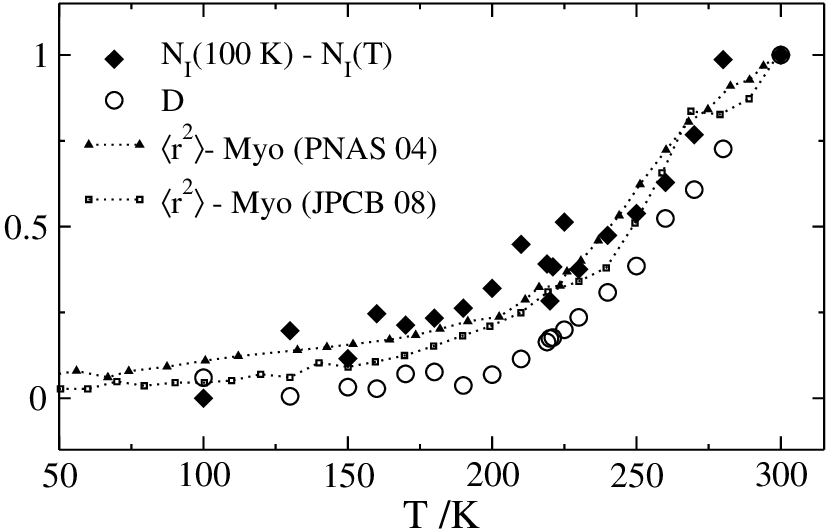}
  \caption{Diffusion coefficient (open circles) from the present
    simulations and atomic mean-squared displacements of myoglobin
    measured experimentally  \cite{Fenimore:04} (small up-triangles)
    and obtained from MD simulations \cite{Krishnan:08} (small
    squares).  Closed diamonds show the change in the number of
    particles in the first solvation shell (Fig.\ \ref{fig:3}). All
    parameters are normalized to their corresponding values at 300 K.
  }
  \label{fig:6}
\end{figure}

We have shown that the dramatic rise of the reorganization energy
correlates with the depletion of the first solvation shell and the
related increase in the strength of the first-shell density
fluctuations. Figure \ref{fig:6} additionally supports this view.
Here we compare experimental \cite{Fenimore:04} and simulated
\cite{Krishnan:08} atomic mean-squared displacements of myoglobin
(small points) with our calculations of the diffusivity of water in
the simulation box and the change in the number of waters in the first
solvation shell. All parameters have been normalized to their
corresponding values at 300 K to bring them to the common scale. The
remarkable result of this comparison is that the average number of
waters in the first solvation shell follows very closely the atomic
displacements changing its temperature slope at the point of dynamical
transition, $T_{\text{tr}}=220$ K. The increased mobility of the
protein is therefore related to the increased translational mobility
of waters \cite{Tarek_PhysRevLett:02,Tournier:03} caused in turn by
the creation of the high-temperature hydrophobic interface
\cite{ChandlerNature:05}.

\begin{figure}
  \centering
  \includegraphics*[width=6.5cm]{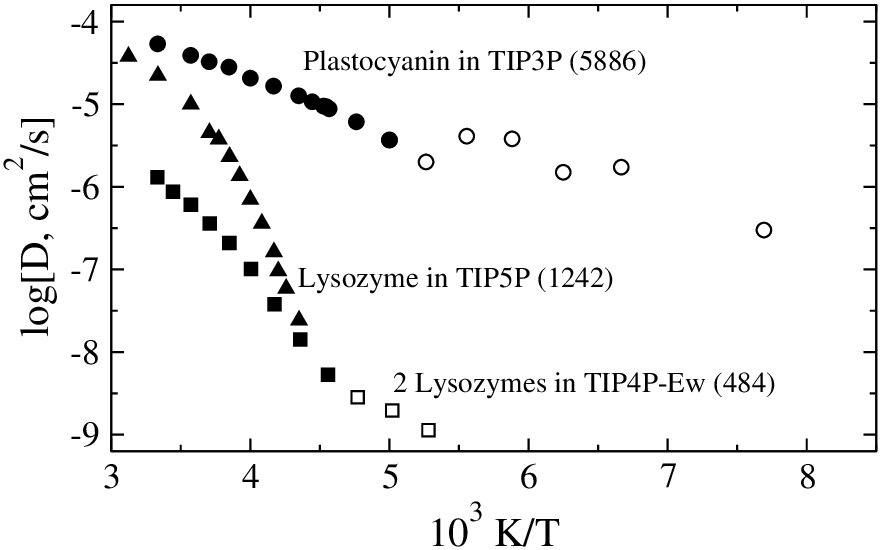}
  \caption{Diffusion coefficient of TIP3P water calculated from all
    $N_w = 5886$ water molecules in the simulation box (circles). 
    The triangles indicate the results of Ref.\ \onlinecite{Kumar:06} 
    for the simulation box containing $N_w=1242$ waters; the squares 
    denote the results from Ref.\ \onlinecite{Lagi:08} with $N_w=484$ 
    per simulation box containing two protein molecules. Closed and
    open points indicate temperatures above and below the dynamical
    transition temperature, respectively.  }
  \label{fig:7}
\end{figure}

The diffusion coefficient of water in the simulation box is plotted
separately vs the inverse temperature in Fig.\ \ref{fig:7}, where we
also compare our results to previous simulations by Kumar \textit{et
  al} \cite{Kumar:06} and by Lagi \textit{et al} \cite{Lagi:08}. The
diffusion coefficient was calculated from the Einstein equation and
the reported values are averaged over all waters in the simulation
box. The different magnitudes of diffusivity compared to previous
reports \cite{Kumar:06,Lagi:08} are related to the different force
fields used, but, more importantly, to the different fractions of
water molecules in the simulation sample. Given that all molecules in
the smallest sample in Fig.\ \ref{fig:7} belong to the interface
\cite{Lagi:08}, it is not that surprising that these data show the
slowest diffusion, in agreement with the common expectation of slower
diffusion of waters in thin interfacial layers
\cite{Sinha:08}. Nevertheless, despite the use of a much larger number
of waters ($N_w=5886$ vs $N_w=484$), we confirm here the existence of
a crossover in the Arrhenius slope of water's diffusion coefficient
observed earlier in Ref.\ \onlinecite{Lagi:08}.

Two observations are relevant in respect to the diffusivity data shown
in Fig.\ \ref{fig:7}. First, the temperature law is Arrhenius both
above and below the transition temperature with the slope decreasing
at lower temperatures, in accord with previous observations
\cite{ChenPNAS:06,Lagi:08}. Second, the transition temperature is
shifted down to 200 K compared to 220 K found in simulations of
partially hydrated proteins in Ref.\ \onlinecite{Lagi:08}.  The first
observation implies that we observe only a change in the character of
a secondary, Arrhenius relaxation, as indeed often seen for electron
transfer in proteins \cite{Parak:80}, instead of a fragile-to-strong
transition.  This fact might be related to the often reported
\cite{Ngai:07} disappearance of $\alpha$ relaxation in confined water most
closely related to our simulation conditions. Since $D(T)$ follows
closely the decrease in the number of hydration-shell waters (Fig.\
\ref{fig:6}) a connection of the break in the slope to a secondary
process produced by collective density fluctuations of the hydration
shell seems a reasonable explanation.  The second feature might imply
that the existence and position of the transition temperature depends
on the fraction of surface waters in the system. While all waters in
the simulation setup in Ref.\ \onlinecite{Lagi:08} belonged to the
surface, only roughly 10\% of waters in our simulations find themselves
in the first solvation shell (Fig.\ \ref{fig:3} and Table
\ref{tab:1}). Likewise, we have obtained a fragile-to-strong crossover
by considering only first-shell fluctuations in Fig.\ \ref{fig:5}, but
it is already washed out for the diffusivity averaged over several
hydration layers.

\begin{figure}
  \centering
  \includegraphics*[width=6.5cm]{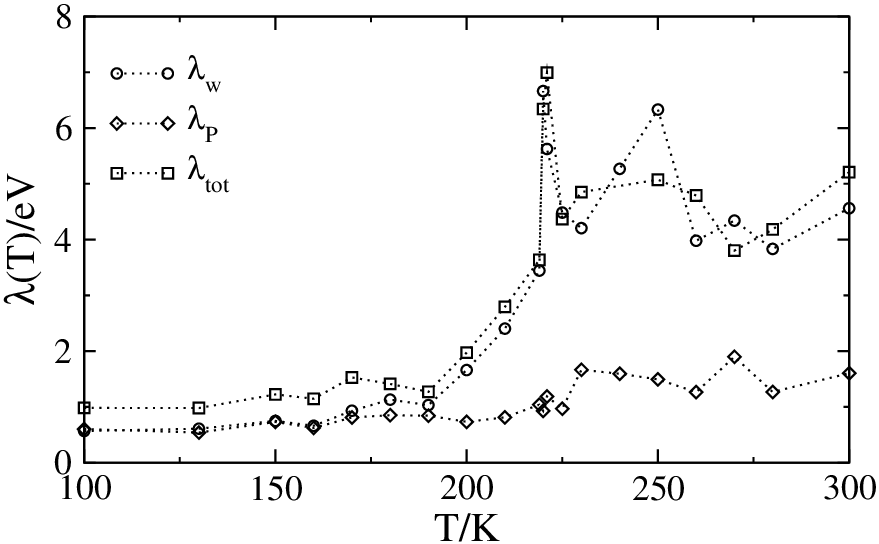}
  \caption{Reorganization energy obtained from the variance of the
    Coulomb energy gap: water component $\lambda_w$ [circles, Eq.\
    (\ref{eq:10})], the protein component $\lambda_P$ from the
    variance of $\Delta E^C_P$ (diamonds) and the total water/protein
    reorganization energy $\lambda_{\text{tot}}$ from the variance of
    the total Coulomb energy gap $\Delta E^C_w + \Delta E_P^C$ (squares).  }
  \label{fig:8}
\end{figure}

Our data, while pointing mostly to the interfacial effects as the
reason for the dramatic rise of $\lambda_w(T)$, do not entirely exclude
crossing the Widom line, a bulk property of water, from the
picture. While the global rise of the intensity of electrostatic
fluctuations within protein is linked to the density fluctuations of
the interface, the spike of $\lambda_w(T)$ at $T=220$ K and the
corresponding slowing down of the Stokes shift relaxation might well
be linked to the crossing of the Widom line.  The increased
cooperativity of water's fluctuations at this temperature causes a
behavior similar to the critical slowing down with a peak in a second
energy cumulant, heat capacity for bulk measurements \cite{Kumar:06}
and reorganization energy in our case.

The spike of $\lambda_w(T)$, barely seen on the 1 ns observation window,
becomes more pronounced on the 10 ns time-scale, as is shown in Fig.\
\ref{fig:8} where the reorganization energies for water, protein, and
the full reorganization energy from water/protein electrostatic
fluctuations were collected from the entire 10 ns trajectories by
calculating the variance of the total Coulomb energy gap $\Delta E_w^C + \Delta
E_P^C$. This observation suggests that some collective motions of
water, significantly cut off on the 1 ns time-scale, contribute to the
peak and become more pronounced on a longer observation scale.

\section{Conclusions}
\label{sec:5}
In conclusion, we have found a dramatic increase in the breadth of
water-induced electrostatic fluctuations inside the protein with
increasing temperature. We link this increase to the creation of the
hydrophobic interface at extended hydrophobic patches of the protein.
What has escaped the attention of all studies of the dynamical
transition in biopolymers is the onset of hydrophobic solvation
occurring at the same temperature $T_{\text{tr}}$ as the dynamical
transition. It might be true that the creation of the hydrophobic
interface with its large extent of density fluctuations and intense
electrostatic noise is closely linked to the dynamical transition,
although we do not currently have any additional data supporting this
view. However, if this view is correct, there should be a critical
polypeptide dimension below which the macroscopic hydrophobic
interface does not form \cite{ChandlerNature:05} and no dynamical
transition exists. In fact, very recent measurements of terahertz
dielectric response \cite{He:08} of hydrated polypeptides of different
lengths have indicated the existence of an exactly such critical
polymer length below which the dynamical transition disappears.

We found that the dynamics of electrostatic fluctuations are coupled
to fast $\beta$ relaxation of the hydration shell. The redox activity of
proteins can therefore be classified as hydration-shell-coupled,
according to the classification suggested by Fenimore \textit{et al}
\cite{Fenimore:04}. Although this coupling carries similarities with
aqueous mixtures of simple glass-formers \cite{Ngai:08}, proteins are
not just large molecules. The formation of the hydrophobic interface
is related to a particular lengthscale of hydrophobic patches ($\simeq 1$
nm \cite{ChandlerNature:05}) which does not exist for small hydrated
molecules. Not surprisingly, large-amplitude electrostatic
fluctuations observed here are not usually seen inside small molecules
\cite{DMacc:07}, although this feature might extend to other patchy
hydrophobic surfaces, such as lipid membranes and dendrimeric
structures.

It remains to be seen whether and how the gigantic reorganization
energy found at high temperatures is related to the biological
function of metalloproteins belonging to energetic electron-transfer
chains. One can anticipate, from a general perspective, that a
significant increase in the amplitude of electrostatic fluctuations
can help in reducing barriers for chemical transformations by allowing
better chances for favorable configurations from a broad fluctuation
spectrum.

\acknowledgments This research was supported by the the NSF (CHE-0616646).

\bibliographystyle{apsrev}
\bibliography{/home/dmitry/p/bib/chem_abbr,/home/dmitry/p/bib/liquids,/home/dmitry/p/bib/protein,/home/dmitry/p/bib/dm,/home/dmitry/p/bib/bioet,/home/dmitry/p/bib/glass,/home/dmitry/p/bib/et,/home/dmitry/p/bib/statmech,/home/dmitry/p/bib/simulations,/home/dmitry/p/bib/solvation,/home/dmitry/p/bib/dynamics}

\end{document}